\documentclass[%
superscriptaddress,
preprint,
showpacs,preprintnumbers,
 amsmath,amssymb,
prl,
]{revtex4-1}

\usepackage{graphicx}
\usepackage{color}
\usepackage{dcolumn}
\usepackage{bm}
\usepackage[colorlinks,linkcolor=blue]{hyperref}
\usepackage{CJK}
\usepackage{upgreek}
\usepackage[symbol*]{footmisc}

\makeatletter
\renewcommand*{\@fnsymbol}[1]{\ensuremath{\ifcase#1\or \dagger\or \ddagger\or *\or
\mathsection\or \mathparagraph\or \|\or **\or \dagger\dagger
\or \ddagger\ddagger \else\@ctrerr\fi}}
\makeatother

\begin{document}
\begin{CJK*}{GBK}{}

\title{Intrinsic Ferromagnetism in Electrenes}
\author{Jun \surname{Zhou} }
\email{phyzjun@nus.edu.sg}
\affiliation{Department of Physics, National University of Singapore, Singapore 117551}

\author{Yuan Ping \surname{Feng} }
\email{phyfyp@nus.edu.sg}
\affiliation{Department of Physics, National University of Singapore, Singapore 117551}
\affiliation{Center for Advanced 2D Materials, National University of Singapore, Singapore 117546}

\author{Lei \surname{Shen}}
\email{shenlei@nus.edu.sg}
\affiliation{Department of Mechanical Engineering, National University of Singapore, Singapore 117575}


\begin{abstract}

We report intrinsic ferromagnetism in monolayer electrides or electrenes, in which excess electrons act as anions. Our first-principles calculations demonstrate that magnetism in such electron-rich two-dimensional (2D) materials originates from the anionic electrons rather than partially filled \textit{d} orbitals, which is fundamentally different from ferromagnetism found in other 2D intrinsic magnetic materials. Taking the honeycomb LaBr$_2$ (La$^{3+}$Br$^{-}_{2}\cdot e^{-}$) as an example, our calculations reveal that the excess electron is localized at the center of the hexagon, which leads to strong Stoner-instability of the associated states at the Fermi energy, resulting in spontaneous magnetization and formation of a local moment. The overlap of extended tails of the wave functions of these electrons mediates a long-range ferromagnetic interaction, contributing to a Curie temperature ($T_\textrm{c}$) of 235 K and a coercive field ($H_\textrm{c}$) of 0.53 T, which can be further enhanced by hole doping. The dual nature, localization and extension, of the electronic states suggests a unique mechanism in such magnetic-element-free electrenes as intrinsic 2D ferromagnets.
\end{abstract}

\maketitle
\end{CJK*}

Electrides are intrinsic electron-rich materials, in which excess electrons, named anionic electrons, are confined in geometrically interstitial sites and act as anions \cite{electride_review_science}. Such anionic electrons are loosely-bounded and enable a spectrum of appealing applications in catalysis \cite{electride_catalyst_1,electride_catalyst_2,electride_catalyst_3,electride_catalyst_4}, ideal electron donor \cite{electride_donor_1, electride_donor_2}, batteries \cite{electride_battery_1, electride_battery_2}, superconductivity \cite{zhang2017electride,zhao2019PRL}, and topological matters \cite{electride_topo_park_2018_prl,electride_topo_hirayama_2018_prx,electride_topo_huang_2018_nl,liu2018intrinsic}. Some \emph{bulk} electrides, such as Gd$_2$C \cite{electride_mag_inoshita_2014_prx}, Yb$_5$Sb$_3$ \cite{mott_insulating_electrides}, Y$_2$C \cite{y2c_fm, electride_mag_park_2017_jacs}, high-pressured electrides of Alkali metals \cite{electride_s-band_fm}, and organic electrides \cite{organic_electride_afm_1,organic_electride_afm_2, organic_electride_afm_fm_transition}, have been reported to be magnetic, due to unpaired electrons \cite{electride_mag_zhang_2014_cm,electride_mag_inoshita_2014_prx,electride_mag_park_2017_jacs}. Recently, a new type of electrides with layered structures is attracting increasing research interest due to its unique two-dimensional (2D) electron gas at the interlayer space \cite{electride_mag_inoshita_2014_prx, electride_2D, electride_ca2n} and its potential to be exfoliated into 2D materials \cite{exf_ca2n_zhao_2014_jacs}. For example, monolayer (ML) Ca$_2$N has been exfoliated experimentally and was found to retain its 2D anionic electron gas \cite{ca2n_exf_druffel_2016_jacs}. Such 2D structural electrides are referred to as electrenes \cite{electrene}.

Meanwhile, ferromagnetism (FM) in 2D materials, especially in ferromagnetic semiconductors, has been a hot research topic, because such materials offer special advantages of high data storage density and easy integration into semiconductor devices \cite{feng2017prospects}. However, 2D ferromagnetic semiconductors are rare both in nature and in laboratory due to the weak long-range ferromagnetic order in low dimensional systems \cite{Mermin_Wagner_1966_prl}. Therefore, finding robust 2D magnetic materials has been a challenge for spintronics based on 2D materials \cite{feng2017prospects}. Over the years, there have been some successes in making 2D ferromagnets {\em i.e.,} via defect engineering \cite{ugeda2010missing, peng2013defect}, by introducing magnetic species into semiconductors \cite{seixas2015atomically, li2017two}, and through the magnetic proximity effect from a magnetic substrate \cite{scharf2017magnetic}. All these approaches are ``extrinsic'', lacking the stability and controllability needed for applications. Recently, Gong {\em et al.} \cite{Cr2Ge2Te6_gong_2017_nature} and Huang {\em et al.} \cite{CrI3_huang_2017_nature} first succeeded in fabricating \emph{intrinsic} 2D ferromagnets, Cr$_2$Ge$_2$Te$_6$ and CrI$_3$, respectively, in 2017. Even though the Curie temperature ($T_{\textrm{c}}$) of both materials is low, these experiments demonstrated the possibility of fabricating robust 2D intrinsic ferromagnets. Quickly, more intrinsic 2D ferromagnets, including Fe$_3$GeTe$_2$ \cite{Fe3GeTe2_Zhang_Yuangbo_2018_nature}, VSe$_2$ \cite{VSe2_bonilla_2018_nat_tech}, CrOBr \cite{CrOBr_sunzm_2018_jacs},CrSeBr \cite{screening_zhaojj_2018_acsami}, CrWGe$_2$Te$_6$ \cite{CrWGe2Te6_KanEJ_2018_jacs}, MnPSe$_3$ \cite{MnPSe3_YangJL_2014_jacs}, CoH$_2$ \cite{wu2018transition}, and CrBr$_3$  \cite{Huang2018PRL} were reported.

It is noted that all these intrinsic 2D magnetic materials contain $3d$ magnetic elements such as Cr, Fe, Mn. The magnetism in these materials can be understood based on known exchange models developed for 3D magnetic materials. For example, formation of local magnetic moments in FM semiconductors Cr$_2$Ge$_2$Te$_6$ and CrI$_3$ is associated with the magnetic element Cr which has  partially filled 3\textit{d} subshells. The strongly localized nature of the 3$d$ states under the crystal field favors spontaneous spin polarization and leads to the formation of local moments. The observed collective magnetism is a result of the long-range indirect coupling between these 3$d$ local moments via the spatially delocalized 5$p$ orbital of Te or I, by the superexchange (or \textit{d-p-d} exchange) interaction within the Goodenough-Kanamori-Anderson (GKA) rules \cite{GKA_anderson_1950_pr}. For ferromagnetic systems with free electrons, such as FM metals Fe$_3$GeTe$_2$ or VSe$_2$, the free carriers can mediate a long-range ferromagnetic coupling between the local moments via the Ruderman-Kittel-Kasuya-Yosida (RKKY) exchange (or \textit{s-d} exchange) or the Zener double-exchange (or \textit{p-d} exchange) interaction \cite{RKKY_ruderman_1954_pr, zener1951interaction}.

Given the enormous interest in both electrenes and 2D intrinsic ferromagnetic materials, it is natural to ask whether an electrene can be magnetic. With such a motivation, we screened more than 6,000 structures in the 2D materials database - 2DMatPedia \cite{zhou2dmatpedia} (\url {http://www.2dmatpedia.org/}) for magnetic electrenes. Our search resulted in 9 magnetic electrenes out of a total of 24 electrenes \cite{zhou2019CM}. In contrast to the 2D intrinsic magnetic materials reported so far, magnetic electrenes are fundamentally different in two important aspects: they contain no magnetic elements and the magnetism originates from the anionic electrons (to be discussed below). Therefore, the intrinsic ferromagnetism in magnetic electrenes is completely of a different origin and cannot be explained based on known exchange models. In this Letter, taking LaBr$_2$ as an example, we unveil the origin of local moment and propose a new exchange interaction for long-range ferromagnetic coupling in intrinsic ferromagnetic electrenes. Based on the understanding of the magnetic mechanism, we further propose a method to increase both the Curie temperature and the coercive field.

Our first-principles calculations indicate that the excess electron in LaBr$_2$ is localized at the center of the hexagon of its honeycomb hexagonal crystal structure and carries a magnetic moment of $1\mu_\textrm{B}$. The wave function shows a dual nature of localization and extension. While the localization of the anionic electron's wave function contributes to the formation of the local magnetic moment, the extended tail of the wave function mediates a long-range magnetic interaction. This long-range ferromagnetic order leads to a Curie temperature of 235 K and a coercive field of 0.53 tesla (T) , which can be increased by 45\% and 36\%, respectively, via a $0.1~e$/unit-cell (u.c.) hole doping, or equivalently, a hole concentration of $\sim  6.7\times 10^{13}~\textrm{cm}^{-2}$, indicating its magnetic stability against the thermal and magnetic perturbation.

\begin{figure}[ptb]
\centering
\includegraphics [width=0.95\textwidth]{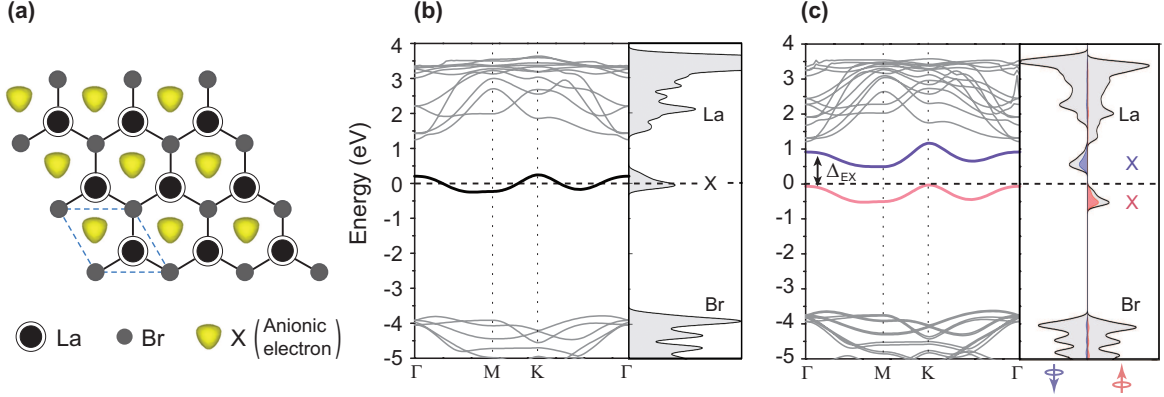}\newline\caption{(a) LaBr$_2$ (La$^{3+}$Br$^{-}_{2}\cdot e^{-}$) ML in H-phase MoS$_2$ structure with ELF maps (isosurface value $= 0.7$) in yellow. The dashed diamond denotes the unit cell. Only the ELF on interstitial sites is shown for clarity. The anionic electron in the geometric space, labeled as X, is at the center of each hexagon. Band structure and density of states of ML LaBr$_2$ (b) without and (c) with spin polarization.}
\label{fig1}%
\end{figure}

The calculations were performed using the density functional theory (DFT) with generalized gradient approximations and all-electron projector augmented wave method implemented in the Vienna \textit{ab initio} simulation package {\footnotesize{(VASP)}} \cite{VASP}. Convergence with respect to the on-site Coulomb interaction, the plane-wave cutoff energy and k-point sampling has been carefully checked. The structures were fully relaxed until the atomic forces were less than 0.001 eV/\AA. Phonon dispersions and \textit{ab initio} molecular dynamics (MD) simulations were carried out to confirm the structural stability. We performed Monte Carlo (MC) simulation with the Metropolis algorithm and micromagnetic simulation using the Landau-Lifshitz-Gilbert (LLG) equation to investigate its magnetic stability against temperature and an external magnetic field [see \textbf{Methodology in Supplemental Material}].

\textit{Structure of monolayer LaBr$_2$.---} Layered bulk LaBr$_2$ was synthesized experimentally in 1989 \cite{Kramer1989}. Its crystal structure resembles that of 2H-phase of MoS$_2$, {\em i.e.,} a honeycomb hexagonal structure with a layer of La atoms sandwiched between two layers of Br atoms (space group P6$_3$/mmc), as shown in \textbf{Fig.~1(a)}. The excess electron in La$^{3+}$Br$^{-}_{2}\cdot e^{-}$ is confined in the center of the hexagon, free from any of the atomic orbitals, and acts as an anion [\textbf{Fig.~1(a)}]. Similar to how graphene is obtained from graphite, layered bulk electrides can be exfoliated into 2D sheets. For example,  ML Ca$_2$N has been experimentally isolated by liquid exfoliation \cite{ca2n_exf_druffel_2016_jacs}. The calculated exfoliation energy of LaBr$_2$ is 0.27 J/m$^2$ which is lower than that of Ca$_2$N (1.13 J/m$^2$)  \cite{exf_ca2n_zhao_2014_jacs} and graphene (0.43 J/m$^2$) \cite{ exf_graphene_zacharia_2004_prb}, indicating that ML LaBr$_2$ can be easily isolated using a similar approach. Furthermore, both the calculated phonon spectrum and \textit{ab initio} molecular dynamics simulation confirm the thermodynamic structural stability of ML LaBr$_2$ [see \textbf{FIG. S4}].

\begin{figure}[ptb]
\centering
\includegraphics [width=0.90\textwidth]{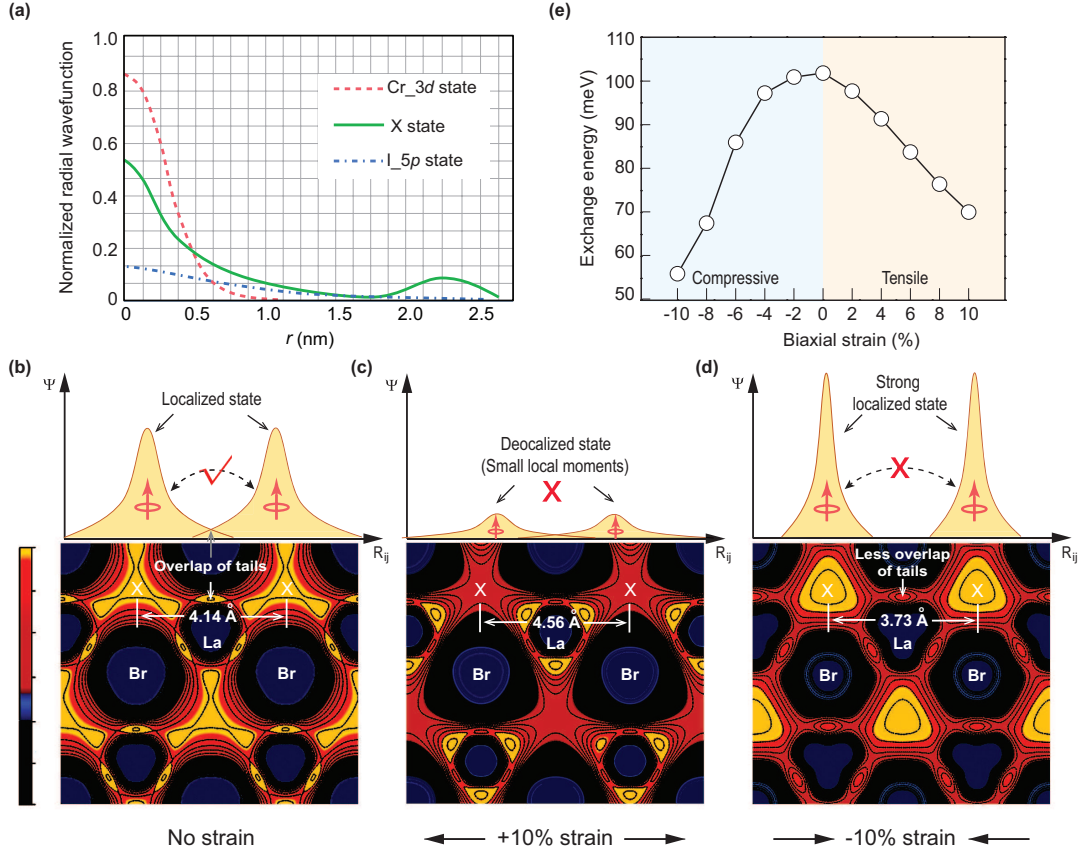}\newline\caption{(a) Radial distributions of normalized wave-function of states of Cr$_{3d}$, I$_{5p}$ and the anionic electron. Spin density of strain-free ML LaBr$_2$ (b), ML LaBr$_2$ with 10$\%$ tensile strain (c), and ML LaBr$_2$ with 10$\%$ compressive strain (d). Schematics of strain induced changes in the localization of wave functions and the overlap of their tails. The isovalue is $0.001 e$/\AA$^3$. (e) The strain dependence of the exchange energy.}%
\label{fig2}%
\end{figure}

\textit{Origin of local moments.---} Two key factors for observed magnetism in magnetic materials are local moments and their long-range coupling \cite{dev2008defect}. In most of the known magnetic materials, the formation of local moment is associated with magnetic elements with partially filled $3d$ subshells \cite{CrI3_huang_2017_nature, Fe3GeTe2_Zhang_Yuangbo_2018_nature}. The strongly localized nature of the 3$d$ states favors spin-polarized electron configurations and leads to the formation of local moments. However, this is not the case in ML LaBr$_2$ which has no magnetic elements. Magnetism in LaBr$_2$ has to be of a complete different origin. \textbf{Figure 1(b)} shows the band structure and density of states (DOS) of ML LaBr$_2$ calculated without spin polarization. As can be seen, there is a localized state at the Fermi level ($E_f$), which is mainly from the interstitial electrons labeled as X in \textbf{Fig.~1(a)}. According to the Stoner criterion, $g(E_f)J>1$, where $g(E_f)$ is the DOS at $E_f$ and $J$ denotes the strength of the exchange interactions \cite{stoner_1938_prsla}, this strongly localized state of the interstitial electron is unstable and would lead to spontaneous spin polarization. The electronic structure of ML LaBr$_2$ with the spin polarization [\textbf{Fig.~1(c)}] shows a local moment of 1$\mu_\textrm{B}$ per unit cell. The spin-exchange splitting energy ($\Delta_{EX}$) is around 1 eV, comparable to that of the Mn$_{3d}$ orbital (1.62 eV) for $M=2~\mu_\textrm{B}$ \cite{peng2009origin}. The calculated electron localization function (ELF), projected DOS, and spin density all confirm that the magnetic moment is mainly contributed by the interstitial electron, labeled as X in \textbf{Fig.~1(a)}.

\textit{Mechanism of long-range magnetic order.---} The magnetic exchange energy is calculated from the energy difference between two magnetic configurations, a ferromagnetic configuration, in which all moments of the anionic electrons are aligned parallel, and an antiferromagnetic (AFM) configuration, in which moments of neighboring rows of the anionic electrons point to opposite directions, in a 2$\times$2 supercell (see \textbf{FIG. S5}). Specifically, the exchange energy is given by $J=(E_\textrm{AFM} - E_\textrm{FM})/4nM^2$, where $E_\textrm{AFM}$, $E_\textrm{FM}$, $n$ and $M$ denotes the total energy of the supercell in the AFM, FM configuration, number of unit cells in the supercell, and local magnetization on the anionic electron, respectively. The calculated positive value of $J$ of 6.38 meV suggests a ferromagnetic order in ML LaBr$_2$. Different Hubbard U values and hybrid functions (HSE) are carefully tested and lead to small differences in $J$ [see \textbf{Supplemental Material}].

It is known that collective magnetization in solids requires long-range ferromagnetic coupling between the local moments of spin-polarized electrons, either itinerant or bound. Examples of the former are the RKKY (or \textit{s-d}) and \textit{p-p} exchange interaction \cite{pan2007room, shen2008mechanism, shen2013simultaneous}, and the latter is the superexchange (or \textit{p-d}) interaction in which the localized $d$ states are mediated by the spatially delocalized $p$ states (see \textbf{FIG. S6}). However, the long-range ferromagnetic order in LaBr$_2$ is neither of these forms and would be of a complete different mechanism since the concerned electronic states are not associated with any atomic orbital. Neither the carrier-mediated \textit{s-d} and \textit{p-p}, nor orbital-mediated \textit{p-d} coupling is applicable to this system. Moreover, the angle of 120 degrees between the anionic electrons and La is not ideal for the ferromagnetic superexchange interaction based on the GKA rules \cite{GKA_anderson_1950_pr}.

To understand the coupling between the moments of the anionic electrons, we  examine the nature of the electronic states.
In \textbf{Fig.~2(a)}, we compare the radial distributions of the wave functions of the Cr$_{3d}$ and I$_{5p}$ orbitals, and the anionic electronic states. It is found that the state of anionic electrons is as localized as the atomic $d$ states, and meanwhile, as extended as the atomic $p$ states. Based on this duality (localization and extension) of the electronic states, the formation of the local moment can be expected from the localized feature, while the long-range magnetic coupling can be attributed to the extended nature of the electronic states. Therefore, it is this dual nature of the state of anionic electrons that is responsible for the ferromagnetism in this electrene.
As shown in \textbf{Fig.~2(a)}, the wave function of the anionic electron state is well localized, similar to the atomic $d$ states. Meanwhile, it has a long tail, similar to the atomic $p$ states. We propose that the long-range ferromagnetic interaction in magnetic-element-free ML LaBr$_2$ is mediated by this extended tail of the wave function, as illustrated in \textbf{Fig.~2(b)}. The calculated spin density of ML LaBr$_2$ in \textbf{Fig.~2(b)} clearly shows overlap of the anionic electron wave functions between neighboring cells. Different from the direct exchange interaction, which favors a short-range AFM order between strongly localized $d$ states, the long extended tails of the anionic electron wave function align the local moments in a long range, as shown in \textbf{Fig.~2(b)} and \textbf{FIG.~S5}.

As the localization and the extension of the anionic electronic states are two necessary conditions for the long-range magnetic order in LaBr$_2$, weakening of either can be expected to reduce the long-range interaction. For example, applying an in-plane biaxial tensile strain of 10$\%$ leads to  delocalization of the wave functions, while an in-plane biaxial compressive strain of 10$\%$ reduces the overlap between the tails of the wave functions, as shown in \textbf{Fig. 2(c)} and \textbf{2(d)}, respectively, both resulting in weaker exchange interaction compared with the undeformed structure, as shown in \textbf{Fig.~2(e)}, and lower Curie temperature.

\begin{figure*}[ptb]
\centering
\includegraphics [width=0.90\textwidth]{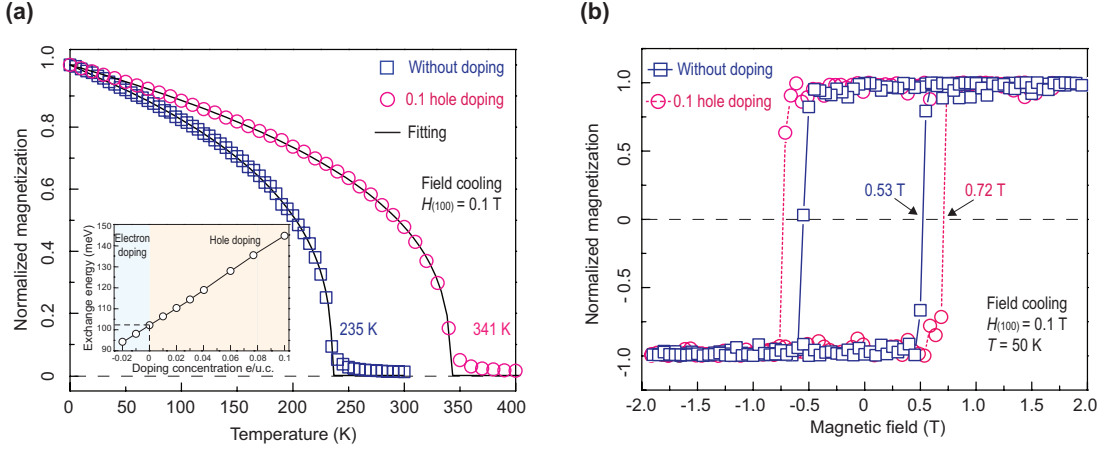}\newline\caption{(a) The temperature-dependence of magnetization without (square) and with (circle) $0.1e$/u.c. doping. The 0.1 T cooling field is applied along the (100) direction. The Inset shows the doping concentration dependence of the exchange energy. (b) The simulated magnetic field-dependence of magnetization at the temperature of 50 K.}%
\label{fig3}%
\end{figure*}

\textit{Magnetic stability and enhancement.---} Curie temperature and the coercive field ($H_\textrm{c}$) are two key parameters for evaluating the magnetic stability of a ferromagnetic material. High $T_\textrm{c}$ and large $H_\textrm{c}$ imply ferromagnetic stability against thermal and magnetic perturbations for practical applications of spintronic devices \cite{Huang2018Naturenanotech., doping_jiang_2018_natnano}. Here, we carry out Monte Carlo simulation with the Metropolis algorithm \cite{Metropolis_binder_1985_jcp,Metropolis_newman_1999_ox} to determine the Curie temperature of LaBr$_2$ and explore possibility of enhancing it.
The spin Hamiltonian of the system is written as
\begin{equation}
H =-J\sum(\textbf{S}_{i}\cdot \textbf{S}_{j})-A\sum(S_{ix}^{2}+S_{iy}^{2}),
\end{equation}
where $\textbf{S}_{i}$ and $\textbf{S}_{j}$ are the classical Heisenberg spins with unit magnitude at the site $i$ and $j$, respectively. $J$ is the nearest neighbor spin-spin coupling, and $A$ is the magnetic anisotropic energy, for which values from the DFT calculations are used. Due to the in-plane anisotropy, the \textit{xy}-type Heisenberg model is adopted with a weak cooling field of 0.1 T along the (100) direction to stabilize the initial magnetic structure according to the Berezinskii-Kosterlitz-Thouless (BKT) transition theory \cite{bkt_berenzinskii_1972_spj, bkt_kosterlitz_1973_jpc}. It is noted that a similar approach was used in a recent experiment on ML Cr$_2$Ge$_2$Te$_6$, in which a cooling field of 0.075 T was applied \cite{Cr2Ge2Te6_gong_2017_nature}. An energy term $H_\textrm{cf}=-g\mu_\textrm{B} H\sum(S_{ix})$, where $g$ is the Land\'{e} g-factor, $\mu_B$ Bohr magneton, $H$ applied external magnetic field (0.1 T), is added to the Hamiltonian in \textbf{Eq.~1}. To validate this model, we calculated $T_\textrm{c}$ of several experimentally reported 2D intrinsic ferromagnets, including both the Ising-type CrI$_3$ and the Heisenberg-type Cr$_2$Ge$_2$Te$_6$ and VSe$_2$. Our results agree with the reported experimental values (see \textbf{Table S4}). The simulated temperature dependent magnetization of ML LaBr$_2$ is shown in \textbf{Fig.~3(a)}, which is fitted to the Curie-Bloch equation in the classical limit $m(T)=(1-T/T_{c})^{\beta}$ \cite{Curie-Bloch_evans_2015_prb}, where $\beta$ is the critical exponent. The calculated $T_\textrm{c}$ of ML LaBr$_2$ is 235 K, in line with 270 K reported in another study \cite{screening_zhaojj_2018_acsami}.

Since many experiments have demonstrated increase of $T_\textrm{c}$ of 2D FM semiconductors by charge doping \cite{doping_jiang_2018_natnano,Huang2018Naturenanotech., Fe3GeTe2_Zhang_Yuangbo_2018_nature}, it is natural to ask whether the $T_\textrm{c}$ of ML LaBr$_2$ can be raised by a similar approach. To answer this question, we repeated calculations at different levels of charge doping. The inset in \textbf{Fig. 3(a)} shows the dependence of the exchange energy on the doping concentration which is varied from $-0.02e$ (electron doping) to $+0.1e$ (hole doping) per unit cell. Our MC simulation shows that merely $0.1e$/u.c. hole doping can augment the $T_\textrm{c}$ to 341 K (45\% enhancement). Charge doping can be easily achieved experimentally by applying an external electric field \cite{doping_jiang_2018_natnano, Huang2018Naturenanotech., Fe3GeTe2_Zhang_Yuangbo_2018_nature}. Here the $0.1e$/u.c. hole doping is equivalent to a hole concentration of $6.7\times 10^{13}~$ cm$^2$, which is comparable to the doping concentration of $2.5\times 10^{13}~$ cm$^2$ in the experiment by Jiang {\em et al.} \cite{doping_jiang_2018_natnano}.

To determine the coercive field ($H_\textrm{c}$) of ML LaBr$_2$, we performed micromagnetic simulation based on the Landau-Lifshitz-Gilbert equation \cite{llg_nakatani_1989_jjap}
\begin{equation}
\frac{\partial \textbf{m}}{\partial t} = -\gamma \mu_s \textbf{m} \times \textbf{H}_\textrm{eff} + \alpha \textbf{m} \times \frac{\partial \textbf{m}}{\partial t},
\end{equation}
where $\textbf{m}$, $\textbf{H}_\textrm{eff}$, $\gamma$, $\mu_s$, and $\alpha$, is the normalized magnetization, applied external magnetic field, gyromagnetic ratio, magnetic moment, and Gilbert damping constant, respectively. The simulation was carried out on a LaBr$_2$ sheet of $1\times 1$ $\mu\textrm{m}^2$, which matches the experimental length scale \cite{Cr2Ge2Te6_gong_2017_nature, CrI3_huang_2017_nature, Fe3GeTe2_Zhang_Yuangbo_2018_nature}. \textbf{Figure 3(b)} shows the magnetic hysteresis loop of ML LaBr$_2$ under a varying external magnetic field. As can be seen, ML LaBr$_2$ has a large coercive field of 0.53 T at 50 K, which is higher than 0.39 T of VS$_2$ \cite{coercive_field_vs2_vatansever_2018_mre, coercive_field_vs2_gao_2013_jmcc} and 0.05 T of CrI$_3$ at the same temperature \cite{doping_jiang_2018_natnano}. Furthermore, a hole doping of $0.1 e$/u.c. not only raises the $T_\textrm{c}$, but also enhances the $H_\textrm{c}$ to 0.72 T (36\% enhancement). This trend is in qualitative agreement with recent experimental observation that hole doping significantly increases both Curie temperature and coercive force of CrI$_3$ \cite{doping_jiang_2018_natnano}.

\textit{Conclusions.---} In conclusion, the origin of the local moment and long-range magnetic order in a magnetic-element-free ML LaBr$_2$ is investigated by first-principles calculations. The state of the anionic electrons is localized but has extended tails. The localized nature of the anionic electron's wave functions favors spin polarization and formation of local moments. The overlap of the extended tails of wave functions, on the other hand, mediates a long-range ferromagnetic order in this electrene. Furthermore, only $0.1e$/u.c. hole doping (or a hole concentration of $6.7\times10^{13}$ cm$^2$) can increase  $T_\textrm{c}$ and $H_\textrm{c}$ by 45\% and 36\%, respectively.

Finally, we wish to point out that we have taken LaBr$_2$, a layered electride, as an example to demonstrate the concept of anionic electron induced long-range ferromagnetic order. However, electrides are a big family \cite{zhou2019CM, electride_mag_inoshita_2014_prx, electride_topo_hirayama_2018_prx, electride_topo_park_2018_prl, sushko2003electron}. The physical mechanism proposed in this Letter can be applied to ferromagnetism in other FM electrides, such as PrI$_2$, YbI$_2$, LaBr and LaCl (see \textbf{FIG. S8}). As reminded by Harry A. Eick thirty years ago \cite{eick1987jlcm}, they are ``still a veritable gold mine" that is worthy further studies on these intrinsic electron-rich materials which may have intriguing properties and unique technological applications. For example, very recently, Liu {\em et al.} reported intrinsic quantum anomalous Hall effect induced by in-plane magnetisation in LaCl \cite{liu2018intrinsic, electride_topo_huang_2018_nl}. Hirayama {\em et al.} proposed a class of electrides with exotic topological properties \cite{electride_topo_hirayama_2018_prx}. We hope this work will stimulate further experimental and theoretical studies to discover more electrenes and explore new spin physics in this unique kind of 2D intrinsic magnetic systems, which may lead to practical spintronic applications.

This work was supported by the National University of
Singapore Academic Research Fund Tier 1 Grant No. R-144-000-361-112, R-265-000-651-114 and R-144-000-413-114.
The Authors thank Dr. Wang Xiao, Dr. Yang Shengyuan, Dr. Yongzheng Luo and Dr. Lei Xu for their helpful discussion. The calculations were carried out on the GRC-NUS and National Supercomputing Centre Singapore high-performance computing facilities.


%

\end{document}